\documentclass[sigconf,review=False,anonymous=False]{acmart}
\usepackage{float}
\usepackage{hyperref}
\usepackage{tikz-qtree}
\usepackage{pifont}
\usepackage[table]{xcolor}
\usepackage[dvipsnames]{xcolor}
\usepackage[normalem]{ulem}
\usepackage[most]{tcolorbox}
\usepackage{fontawesome5}
\usepackage{xcolor}
\usepackage{subcaption}
\usepackage{multirow}
\usepackage{threeparttable}

\newcommand{\redcirc}[1]{%
  \tikz[baseline=(char.base)]{
    \node[shape=circle, draw=none, fill=blue!70!black,
          inner sep=1.2pt, minimum size=4mm,
          text=white, font=\bfseries\footnotesize] (char) {#1};
  }%
}

\acmConference[MSR 2026]{MSR '26: Proceedings of the 23rd International Conference on Mining Software Repositories}{April 2026}{Rio de Janeiro, Brazil}
\usepackage{graphicx} 

\title{Humans Integrate, Agents Fix: How Agent-Authored Pull Requests Are Referenced in Practice}
\author{Islem Khemissi\footnotemark[1], 
Moataz Chouchen\footnotemark[1], 
Dong Wang\footnotemark[2], 
Raula Gaikovina Kula\footnotemark[3]}
\affiliation{
    \institution{\footnotemark[1]Concordia University, \footnotemark[2]Tianjin University, \footnotemark[3]The University of Osaka\\
    \country{\footnotemark[1]Canada, \footnotemark[2]China, \footnotemark[3]Japan}}
    }
    
\email{khemissi.islem@outlook.com, 
moataz.chouchen@concordia.ca, 
dong_w@tju.edu.cn, 
raula-k@ist.osaka-u.ac.jp}
\date{November 2025}

\copyrightyear{2026}
\acmYear{2026}
\setcopyright{cc}
\setcctype{by}
\acmConference[MSR '26]{23rd International Conference on Mining Software Repositories}{April 13--14, 2026}{Rio de Janeiro, Brazil}
\acmBooktitle{23rd International Conference on Mining Software Repositories (MSR '26), April 13--14, 2026, Rio de Janeiro, Brazil}
\acmPrice{}
\acmDOI{10.1145/3793302.3793595}
\acmISBN{979-8-4007-2474-9/2026/04}

\begin{document}

\begin{abstract}
Although coding agents have introduced new coordination dynamics in collaborative software development, detailed interactions in practice remain underexplored, especially for the code review process. In this study, we mine agent-authored PR references from the AIDev dataset \cite{li2025rise} and introduce a taxonomy to characterize the intent of these references across Human-to-Agent and Agent-to-Agent interactions in the form of Pull Requests (i.e. PRs). Our analysis shows that while humans initiate most references to agent-authored PRs, a substantial portion of these interactions are AI-assisted, indicating the emergence of meta-collaborative workflows, where humans mostly use references to build new features, whereas agents make them to fix errors.

We further find that referencing/referenced PRs are associated with substantially longer lifespans and review times compared to isolated PRs, suggesting higher coordination or integration effort. We then list three key takeaways as potential future research directions into how to utilize these dynamics for optimizing AI coding agents in the code review process.

\end{abstract}
\keywords{Code Review, Collaborative software engineering, Mining software repository, Agentic Software engineering}

\maketitle

\section{Introduction}
AI coding agents such as Claude Code, Cursor, Devin, GitHub Copilot, and OpenAI Codex are transitioning from developer assistants to autonomous teammates in the software development lifecycle \cite{li2025rise, zhang2023practices}. These agents can independently generate code, fix bugs, and submit pull requests (PRs), fundamentally altering the dynamics of software engineering \cite{hassan2024rethinking}. As these agent-authored PRs become more prevalent, it is crucial to study how they are integrated into existing development practices, in order to (1) ensure effective collaboration on codebases and (2) assess their impact on the code review process. 

Code review is a cornerstone of modern software development, serving to improve code quality, facilitate knowledge transfer, and ensure adherence to project standards \cite{bacchelli2013expectations, mcintosh2016empirical}. Within this process, the practice of referencing (linking PRs to other PRs, issues, or commits) is vital for establishing shared awareness, providing context, and ensuring traceability \cite{chopra2021alex, zampetti2017developers}. Studying these references provides a unique lens through which to observe the collaborative dynamics and information needs of a project \cite{pascarella2018information}.

\begin{figure}[!ht]
    \centering
    \begin{subfigure}[t]{\linewidth}
        \centering
        \includegraphics[width=\linewidth]{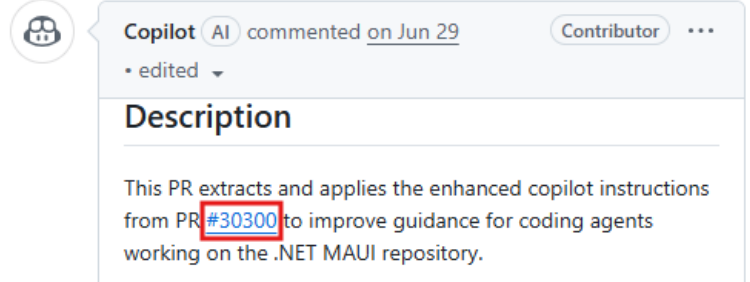}
        \caption{Referencing PR}
        \label{fig:referencing_pr_example}
    \end{subfigure}

    \begin{subfigure}[t]{\linewidth}
        \centering
        \includegraphics[width=\linewidth]{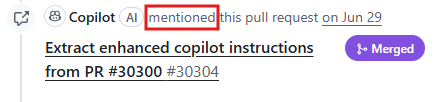}
        \caption{Referenced PR}
        \label{fig:referenced_pr_example}
    \end{subfigure}

    \caption{Example of PR referencing. A reference from one PR creates an event on the \textbf{referenced} PR's timeline (PR \#30300).}
    \label{fig:pr_referencing_example}
\end{figure}

We adopt the terms "\textit{referenced}"/"\textit{referencing}" because they correspond directly to the event name used in the GitHub API, ensuring consistency between our terminology and the underlying data source. In \autoref{fig:referencing_pr_example}, the Copilot agent created a PR (\#30304)\footnote{https://github.com/dotnet/maui/pull/30304} that is \textbf{referencing} another PR (\#30300)\footnote{https://github.com/dotnet/maui/pull/30300}.

With the rise of agent-authored PRs, the nature of these interactions is changing. Are they treated as isolated contributions or as integral parts of the development history? This paper addresses these questions through an empirical study of the AIDev dataset, a large publicly available collection of agent-authored PRs \cite{li2025rise}. Our study is guided by three research questions:
\begin{itemize}
    \item[\textbf{RQ1}] \textbf{ (PR linkage prevalence) To what degree are agent-authored PRs linked in the code review process?}

\end{itemize}
    \noindent In particular, we investigate the prevalence of references to agent-authored PRs, the actors involved in these references (humans vs. agents), and the patterns of self-referencing versus cross-agent referencing.


\begin{itemize}
    \item[\textbf{RQ2}] \textbf{ (Linked vs. Non-Linked PRs) What are the impacts of referencing agent-authored PRs on code review process?}

\end{itemize}

\noindent We analyse the association referencing events in agent-authored PRs and their relation with code review effort, measured through the number of commits (\#Commits), the number of comments (\#Comments) and review duration \cite{wang2021automatic,hirao2019review,alomar2022code}.

\begin{itemize}
    \item[\textbf{RQ3}] \textbf{(Linkage reasons) What are the primary drivers of interaction with agent-authored PRs?}

\end{itemize}

\noindent We develop a taxonomy of referencing intents to classify the motivations behind agent-to-agent and human-to-agent references \cite{hirao2019review,wang2021automatic}.

Our results show that referencing between agent-authored pull requests occurs in 4.2\% of cases, a rate comparable to that observed in large open-source systems such as LibreOffice and Android~\cite{wang2021automatic}. Although most references are initiated by humans, with only about 4\% created by agents, humans frequently rely on agent-generated pull requests when establishing references. We further observe that referenced pull requests are associated with higher review effort. Finally, our analysis of referencing behavior reveals distinct drivers: agents primarily reference pull requests to fix or correct issues, whereas humans more often reference agent-authored pull requests to extend or build upon existing code.

\textbf{Open science}: Our data, scripts, and manual labeling are available in our replication package \cite{replicationpackage}. 

\section{Study Design}
\begin{figure}[!ht]
    \centering
    \includegraphics[width=1\linewidth]{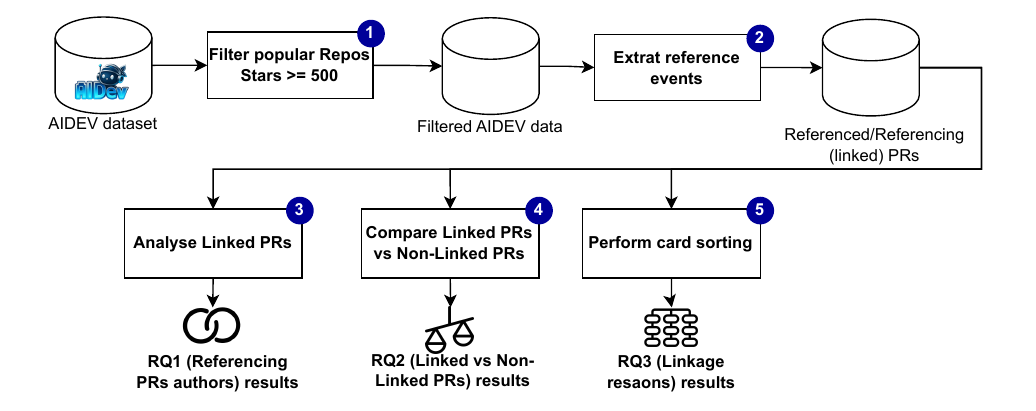}
    \caption{Methodology}
    \label{fig:methodology}
\end{figure}

Figure \ref{fig:methodology} outlines the methodology for our study. 

Our study is based on the AIDev dataset \cite{li2025rise}, which contains 932,791 agent-authored PRs from 116,211 repositories, involving 72,189 developers and five AI coding agents: Claude Code, Cursor, Devin, GitHub Copilot, and OpenAI Codex. In step \redcirc{1}, we filter a subset of 33,596 agent-authored PRs from 2,807 popular repositories (over five hundred stars on Github), which are enriched with detailed metadata. Subsequently, in Step \redcirc{2}, we extract reference events from the PR timelines. In step \redcirc{3}, we categorize these references as follows:
\begin{itemize}
    \item \textbf{Agent-to-Agent references}: References made by agent-authored PRs or commits to other agent-authored PRs.
        \begin{itemize}
            \item \textbf{Self-reference}: References made by agent-authored PRs to PRs authored by the same agent.
            \item \textbf{Cross-agent reference}: References made by agent-authored PRs to agent-authored PRs authored by different agents.
        \end{itemize}
    \item \textbf{Human-to-Agent references}: References made by human-authored PRs or commits to agent-authored PRs.
\end{itemize}

\noindent Furthermore, for human-to-agent references, we distinguished between \textbf{Solo Human} references and \textbf{AI-Assisted Human} references by examining the co-authorship information from the commit description available in the dataset. This allowed us to quantify the extent to which humans use AI tools to help them interact with agent-authored PRs. This analysis helps us answer RQ1. In Step \redcirc{4}, we compare linked PRs (Referenced or Referencing PRs) with PRs without any linkage to understand the impact of linkage on reviewing linked agent-authored PRs. In the comparison, we focused on characterizing review/integration effort measured in terms of the number of PR commits (\#Commits), the number of comments (\#Comments), and the review duration in hours (Duration). The choice of these metrics is motivated by previous studies \cite{wang2021automatic, hirao2019review,wang2021large,alomar2022code,huang2022reviewing,chouchen2023learning,harbaoui2024impact,chouchen2024software,kansab2025empirical}. We apply the Mann–Whitney U test~\cite{conover1999practical}, as PR and code review data are known to be highly skewed~\cite{alomar2022code,coelho2021empirical}. The null hypothesis assumes no difference between PRs with/without linkage, while the alternative hypothesis indicates a significant difference in the measured metrics. We further quantify the magnitude of observed differences using Cliff’s delta ($\delta$) effect size~\cite{cliff1993dominance}, interpreted as negligible for $|\delta| < 0.147$, small for $0.147 \leq |\delta| < 0.33$, medium for $0.33 \leq |\delta| < 0.474$, and large for $|\delta| \geq 0.474$~\cite{romano2006appropriate}. The results of this analysis will help us answer RQ2. 

To answer RQ3, we perform in Step \redcirc{5} card sorting \cite{wood2008card} similar to the work of Hirao et al \cite{hirao2019review}. We manually labeled a total of 213 reference events. Given the limited volume of \textbf{Agent-to-Agent} references, we analyzed the entire population of \textbf{88 instances}. For Human-to-Agent references, we employed stratified sampling to select 25 instances for each of the five agents (totaling \textbf{125 instances}). This strategy ensures equal representation across agents and maintain a dataset size comparable to the Agent-to-Agent set, facilitating a balanced comparison. The first and second authors jointly investigated the reason for linkage, refining the taxonomy through multiple iterations.

\section{Experimental Results}
This section presents the findings of our empirical study.
\subsection{RQ1: To what degree are agent-authored PRs linked in the code review process?}

Our analysis reveals that agent-authored PRs are frequently referenced.

\begin{figure}[!ht]
    \centering
    \includegraphics[width=1\linewidth]{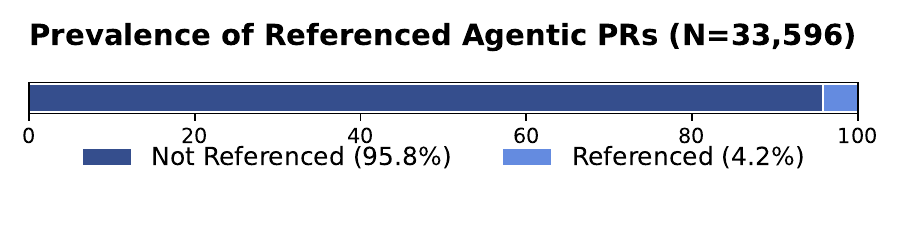}
    \caption{Agent-authored PRs Reference Prevalence}
    \label{fig:rq1_0_prevalence_of_references}
\end{figure}

\textbf{Finding \#1.1: 4.2\% of agent-authored PRs are referenced}. As shown in \autoref{fig:rq1_0_prevalence_of_references}, out of 33,596 agent-authored PRs, 1,412 unique PRs are referenced, corresponding to a linkage rate of 4.20\%. This linkage rate is lower than highly collaborative human communities observed by Hirao et al. \cite{hirao2019review} like Openstack (25\%) or Chromium (17\%) and remains comparable to projects like Android (5\%) and LibreOffice (3\%) \cite{hirao2019review}.

Notably, these 1,412 PRs account for a \textbf{total of 2,010 reference events}, indicating that a subset of agent-authored PRs is referenced multiple times\footnote{https://github.com/cartography-cncf/cartography/pull/1614}. 

\textbf{Finding \#1.2: 95.62\% of total references are initiated by Humans}. As shown in \autoref{tab:interaction_types}, the vast majority of references to agent-authored PRs are made by human-authored changes (commits/PRs). Out of 2,010 total references analyzed, \textbf{1,922 (95.62\%)} were from humans, and only \textbf{88 (4.38\%)} were from agents.

\begin{table}[!ht]
    \centering
    \caption{Distribution of Reference Interactions by Actor}
    \label{tab:interaction_types}
    \resizebox{\columnwidth}{!}{%
        \begin{tabular}{lllcc}
            \toprule
            \textbf{Interaction Type} & \textbf{Source (Actor)} & \textbf{Target (Recipient)} & \textbf{Count} & \textbf{Dist.} \\
            \midrule
            Human-to-Agent & Human Developer & AI Agent & 1,922 & 95.6\% \\
            Agent-to-Agent & AI Agent & AI Agent & 88 & 4.4\% \\
            \bottomrule
        \end{tabular}%
    }
\end{table}

\begin{figure}[!ht]
    \centering
    \includegraphics[width=1\linewidth]{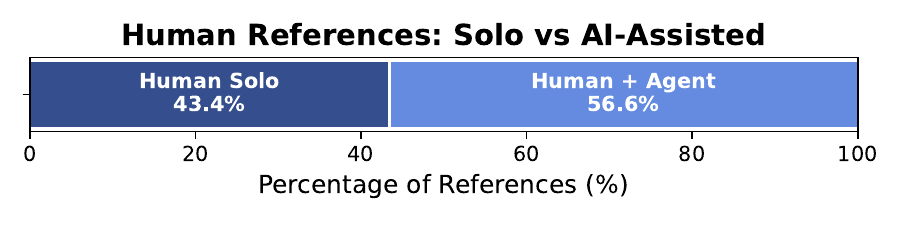}
    \caption{Human-Agent References: Solo Human vs AI-Assisted Distribution}
\label{fig:rq1_4_human_agent_usage_distribution_global}
\end{figure}

\textbf{Finding \#1.3: 56.6\% of Human references are AI-assisted PRs}. Human references to agent-authored PRs are mostly made with AI assistance (43.4\%) as shown in \autoref{fig:rq1_4_human_agent_usage_distribution_global}.

This means that humans are frequently using agents not just to generate code, but also to help them interact with and reference other (or same) agents' work. This "meta-collaboration" pattern suggests a more sophisticated integration of AI into development workflows than previously recognized.



\textbf{Finding \#1.4: Only 4.5\% of agents reference other agents}. When agents do reference other PRs, they rarely reference other agents' PRs. As detailed in \autoref{tab:agent_agent_breakdown}, \textbf{95.45\%} of agent-to-agent references are self-references. This suggests that agents are more often used to build upon their own previous work rather than to collaborate with or correct the work of other agents.
\begin{table}[ht]
    \centering
    \caption{Breakdown of Agent-to-Agent References (N=88)}
    \label{tab:agent_agent_breakdown}
    \resizebox{\columnwidth}{!}{%
        \begin{tabular}{llcc}
            \toprule
            \textbf{Reference Behavior} & \textbf{Description} & \textbf{Count} & \textbf{Dist.} \\
            \midrule
            Self-Referencing & Agent references its own PR & 84 & 95.5\% \\
            Cross-Referencing & Agent references another Agent's PR & 4 & 4.5\% \\
            \bottomrule
        \end{tabular}%
    }
\end{table}

\subsection{RQ2: What are the impacts of referencing agent-authored PRs on the code review process?}
\textbf{Finding \#2.1: Referencing Behavior Follows a Bimodal Pattern}. The timing of references to agent-authored PRs is not uniform. As shown in \autoref{tab:temporal_fingerprint}, references tend to be created either very shortly after the original PR (within an hour) or after a significant delay (more than a week). This bimodal distribution suggests that references serve two primary purposes: immediate and in-context action during the active development, and long-term references for recurring issues or patterns. 

\begin{table}[ht]
    \centering
    \small 
    \setlength{\tabcolsep}{3pt} 
    \caption{Temporal Fingerprint: How Fast Do Agents Link?}
    \label{tab:temporal_fingerprint}
    \begin{tabular}{lccccc}
        \toprule
        & \multicolumn{5}{c}{\textbf{Time Elapsed}} \\
        \cmidrule(lr){2-6}
        \textbf{Agent} & \textbf{$<$1h} & \textbf{1h--1d} & \textbf{1d--1w} & \textbf{1w--1m} & \textbf{$>$1m} \\
        \midrule
        Claude\_Code   & 75\% & 25\% & -- & -- & -- \\
        Devin          & 96\% & 4\%  & -- & -- & -- \\
        OpenAI\_Codex  & 100\%& --   & -- & -- & -- \\
        Cursor         & --   & --   & -- & 100\% & -- \\
        \bottomrule
    \end{tabular}
\end{table}

\textbf{Finding \# 2.2: Linked PRs require more review/integration efforts than compared to non-linked PRs}.
When an agent-authored PR is "\textit{Linked}" (meaning it either references another PR or is referenced by one), its characteristics (\autoref{tab:side_by_side_stats}) change compared to "\textit{Non-linked}" PRs (neither referenced nor referencing other PRs). The results reveal that linked PRs take longer to review in agent-generated pull requests. Our data shows that linked PRs contain double the number of commits compared to isolated ones and involve more comments and discussion. For instance, we observed that \texttt{PR\#868}\footnote{https://github.com/go-vikunja/vikunja/pull/868} took more than 20 days to be closed.

    

\begin{table}[ht]
    \centering
    \fontsize{7}{8}\selectfont
    \tabcolsep=0.04cm
    \caption{\#Commits, \#Comments and Review duration for Linked vs. Non-Linked PRs.}
    \label{tab:side_by_side_stats}

    \begin{threeparttable}
    \resizebox{\columnwidth}{!}{%
        \begin{tabular}{lcccccccccccccc}
            \toprule
            \multirow{2}{*}{\textbf{Metric}} 
            & \multicolumn{2}{c}{\textbf{Min}} 
            & \multicolumn{2}{c}{\textbf{Q1}} 
            & \multicolumn{2}{c}{\textbf{Mean}} 
            & \multicolumn{2}{c}{\textbf{Median}} 
            & \multicolumn{2}{c}{\textbf{Q3}} 
            & \multicolumn{2}{c}{\textbf{Max}} 
            & \multicolumn{2}{c}{\textbf{Stats}} \\
            \cmidrule(lr){2-3} \cmidrule(lr){4-5} \cmidrule(lr){6-7}
            \cmidrule(lr){8-9} \cmidrule(lr){10-11} \cmidrule(lr){12-13}
            \cmidrule(lr){14-15}
             & \textbf{N-Lnk} & \textbf{Lnk}
             & \textbf{N-Lnk} & \textbf{Lnk}
             & \textbf{N-Lnk} & \textbf{Lnk}
             & \textbf{N-Lnk} & \textbf{Lnk}
             & \textbf{N-Lnk} & \textbf{Lnk}
             & \textbf{N-Lnk} & \textbf{Lnk}
             & \textbf{$p$-val} & \textbf{$\delta$} \\
            \midrule
            
            \#Commits
            & 0 & \textbf{1} & 1 & 1 & 1.9 & \textbf{4.2}
            & 1 & \textbf{2} & 2 & \textbf{5} & \textbf{27} & 25
            & $<.01$ & 0.41 (M) \\
            
            \#Comments
            & 0 & 0 & 0 & \textbf{1} & 0.8 & \textbf{1.9}
            & 0 & \textbf{2} & 1 & \textbf{3} & \textbf{14} & 6
            & $<.01$ & 0.56 (L) \\
            
            Review duration (h)
            & 0 & 0 & 0 & \textbf{0.3} & \textbf{32.4} & 32.3
            & 0.1 & \textbf{1.3} & 2.7 & \textbf{18.3}
            & \textbf{1778} & 1148
            & $<.01$ & 0.37 (M) \\
            
            \bottomrule
        \end{tabular}
    }
    \begin{tablenotes}
        \footnotesize
        \item Lnk: Linked PR, N-Lnk: Non-Linked PR
    \end{tablenotes}
    \end{threeparttable}
\end{table}

\subsection{RQ3: What are the primary drivers of interaction with agent-authored PRs?}

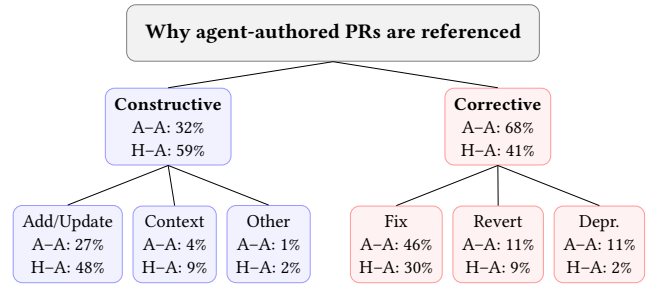
\begin{figure}[!ht]
\centering
\resizebox{\columnwidth}{!}{%
\begin{tikzpicture}[
    grow=down,
    level distance=1.8cm,
    edge from parent/.style={
        draw,
        edge from parent path={(\tikzparentnode.south) -- (\tikzchildnode.north)}
    },
    root/.style={
        draw=black!50, fill=gray!10,
        rounded corners, align=center,
        font=\bfseries, inner sep=8pt
    },
    cons/.style={
        draw=blue!40, fill=blue!5,
        rounded corners, align=center,
        font=\small, inner sep=4pt
    },
    corr/.style={
        draw=red!40, fill=red!5,
        rounded corners, align=center,
        font=\small, inner sep=4pt
    },
    level 1/.style={sibling distance=0.6cm},
    level 2/.style={sibling distance=0.15cm}
]

\Tree
[.\node[root]{Why agent-authored PRs are referenced};
    [.\node[cons]{\textbf{Constructive}\\
        A--A: 32\%\\
        H--A: 59\%
    };
        [.\node[cons]{Add/Update\\
            A--A: 27\%\\
            H--A: 48\%
        }; ]
        [.\node[cons]{Context\\
            A--A: 4\%\\
            H--A: 9\%
        }; ]
        [.\node[cons]{Other\\
            A--A: 1\%\\
            H--A: 2\%
        }; ]
    ]
    [.\node[corr]{\textbf{Corrective}\\
        A--A: 68\%\\
        H--A: 41\%
    };
        [.\node[corr]{Fix\\
            A--A: 46\%\\
            H--A: 30\%
        }; ]
        [.\node[corr]{Revert\\
            A--A: 11\%\\
            H--A: 9\%
        }; ]
        [.\node[corr]{Depr.\\
            A--A: 11\%\\
            H--A: 2\%
        }; ]
    ]
]
\end{tikzpicture}
}
\caption{Unified taxonomy of pull-request reference types. Percentages indicate the distribution of Agent--Agent (A--A) and Human--Agent (H--A) interactions within each category.}
\label{fig:unified_taxonomy_merged}
\end{figure}

The \autoref{fig:unified_taxonomy_merged} summarizes why agent-authored PRs are referenced, distinguishing between two interaction types:
\begin{itemize}
    \item \textbf{A-A}: Agent-to-Agent references
    \item \textbf{H-A}: Human-to-Agent references
\end{itemize}

\textbf{Finding \#3.1: A-A and H-A references reveal two types of referencing behaviour: Constructive and Corrective referencing}

\noindent After performing card sorting on the reasons of linkage in agent-authored PRs, we explore the following linkage reasons behaviours: 

\textcolor{gray}{ \faTags\ }
\textbf{Constructive references} extend the referenced work by adding new features, updating existing logic, or providing necessary context, accounting for \textbf{59\%} of H-A references compared to only \textbf{32\%} for A-A interactions.

\textbf{Add / Update}: 
Humans are nearly twice as likely (\textbf{48\%}) as agents (\textbf{27\%}) to reference agent-authored PRs in order to add new functionality or update existing logic.

This reinforces the role of humans as primary integrators, using agent-generated code as a foundation for further development.
Agents, by contrast, show limited additive behavior, suggesting that they are rarely deployed to incrementally extend previous agent-authored contributions.

In \texttt{PR\#1842}\footnote{https://github.com/imbhargav5/rooks/pull/1842}, a human developer added unit tests and updated the README file based on \texttt{PR\#1841}\footnote{https://github.com/imbhargav5/rooks/pull/1841}.

\textbf{Context}: 
Contextual references (such as linking related PRs for explanation or traceability) are rare overall, but more common in H-A interactions (\textbf{9\%}) compared to A-A (\textbf{4\%}). 
This indicates that humans are more likely to explicitly manage project knowledge and historical context, while agents focus on code-level changes.

For example, \texttt{PR\#53238}\footnote{https://github.com/airbytehq/airbyte/pull/53238} was referenced by \texttt{PR\#53606}\footnote{https://github.com/airbytehq/airbyte/pull/53606} to use the existing "version check logic" code from it. 

\textbf{Other}: 
These low-frequency references (1\% and 2\% for A-A and H-A respectively) reflect edge cases that do not clearly fit into the main taxonomy and do not significantly influence the overall interaction patterns. We found only one instance for backporting where \texttt{PR\#10478}\footnote{https://github.com/dotnet/aspire/pull/10478} was created to backport \texttt{PR\#10453}\footnote{https://github.com/dotnet/aspire/pull/10453}.


\textcolor{gray}{ \faTags\ }
\textbf{Corrective references} aim to remediate issues by fixing bugs, reverting changes, or deprecating the referenced code. These actions are performed by agents (\textbf{68\%}) more than human (\textbf{41\%}).



\textbf{Fix}: 
Fixing is the most common corrective action in both interaction types, but it is more prevalent in A-A (\textbf{46\%}) references compared to H-A (\textbf{30\%}). This suggests that agents are frequently used to repair defects introduced by earlier agent-generated PRs.

We found that \texttt{PR\#810}\footnote{https://github.com/pdfme/pdfme/pull/810} referenced its previous \texttt{PR\#809}\footnote{https://github.com/pdfme/pdfme/pull/809} to fix the "type conversion error".

\textbf{Revert}: 
Revert actions occur at similar rates across interaction types (11\% for A-A vs. 9\% for H-A), indicating that both humans and agents sometimes conclude that agent-generated changes should be undone entirely.
This reflects limits in initial solution validity or contextual appropriateness.

In the following example, \texttt{PR\#2127}\footnote{https://github.com/open-ani/animeko/pull/2127} reverted the changes that were made in \texttt{PR\#2135}\footnote{https://github.com/open-ani/animeko/pull/2135} that caused an error during the application installation.

\textbf{Deprecate}: 
Deprecation is more common in A-A (\textbf{11\%}) interactions compared to H-A (\textbf{2\%}). This suggests that agents may generate follow-up PRs to explicitly retire their own prior changes, while humans rarely engage in formal deprecation when interacting with agent-authored PRs.

For instance, one of the commits made in \texttt{PR\#13269}\footnote{https://github.com/near/nearcore/pull/13269} referenced \texttt{PR\#13267}\footnote{https://github.com/near/nearcore/pull/13267} to deprecate some of its functions (e.g. IncreaseDeploymentCost, LimitContractLocals, LowerStorageKeyLimit).


In our dataset, Cursor's agent-to-agent self-references are exclusively (100\%) corrective, while Claude Code's are exclusively (100\%) constructive, suggesting agent-specific interaction profiles. Other agents like Copilot and Devin show a more mixed profile.

\section{Takeaways and Future Agenda}

We presented a large-scale empirical study of 33,596 agent-authored pull requests to characterize how they are referenced and integrated into code review workflows. Our analysis reveals that while agent-authored PRs are referenced less frequently than human PRs, they are deeply embedded in human-AI collaborative workflows. Our key takeaways are:

\begin{itemize}
    \item \textbf{Humans integrate, agents self-correct}: We observe two distinct collaboration models based on referencing intent. Humans primarily reference agent-authored PRs to build upon them (59\% constructive), acting as integrators who extend the agent's work. In contrast, agents primarily reference their own prior work to fix it (68\% corrective), engaging in a loop of iterative self-refinement.
    \item \textbf{Linkage is associated with longer PR lifespans and review times}: Linked PRs, which represent only 4.2\% of the dataset, are not trivial contributions. They are associated with higher work and discussion effort and take substantially longer to reach a decision compared to their isolated counterparts. This indicates that when agent-authored PRs are integrated into code review process, they require significantly more human attention.
    \item \textbf{A new form of human-AI teaming is emerging}: The finding that 43.4\% of human references are AI-assisted reveals a "\textit{meta-collaboration}" pattern. Developers are not just using agents to generate code, but also to help them understand, critique, and build upon other AI-generated artifacts, signaling a deeper integration of AI into the entire development lifecycle.
\end{itemize}

For future work, these findings suggest that the integration of agent-authored PRs is not a monolithic process. Instead, it involves a combination of agent specialization, task complexity, and distinct human-AI collaboration patterns. Future work should continue to explore these dynamics to build more effective software development tools.

\bibliographystyle{ACM-Reference-Format}
\bibliography{references.bib}

\end{document}